\newif
\ifconfver
\confvertrue        
\ifconfver
    \documentclass[10pt,journal,final,twoside]{IEEEtran}
\else
    \documentclass[11pt,draftcls,onecolumn]{IEEEtran}
\fi

\usepackage{xspace,empheq,fancybox,amssymb,amsfonts,graphicx,epstopdf,epsfig,syntonly,times,subfigure} 
\usepackage{caption}
\usepackage{psfrag,color,bm,array,tabularx}
\usepackage{cite,url,footnote,xspace,syntonly,algorithmic}
\usepackage{verbatim,multirow}
\usepackage[linesnumbered,ruled,vlined]{algorithm2e}
\usepackage{nomencl}
\makenomenclature

\usepackage{textcomp}
\usepackage{xcolor}
\usepackage{mathtools}
\usepackage{bbm, booktabs}
\usepackage{amsmath, amsthm, lipsum}       

\newcolumntype{M}[1]{>{\centering\arraybackslash}m{#1}}
\newcolumntype{N}{@{}m{0pt}@{}}

\DeclareMathOperator
*{\argmax}{\arg\max}

\def\BibTeX{{\rm B\kern-.05em{\sc i\kern-.025em b}\kern-.08em
		T\kern-.1667em\lower.7ex\hbox{E}\kern-.125emX}}

\newtheorem{remark}{\bfseries Remark}
\input{mysymbol.sty}
\allowdisplaybreaks

\newcommand{\KBC}{\color{black}{}}
\newcommand{\KBD}{\color{black}{}}
\IEEEoverridecommandlockouts
\usepackage{ifthen}
  \renewcommand{\nomgroup}[1]{%
  \item[\bfseries
  \ifthenelse{\equal{#1}{A}}{Indices}{%
  \ifthenelse{\equal{#1}{C}}{Parameters and Variables}{%
  \ifthenelse{\equal{#1}{D}}{Vectors and Matrices}{%
  \ifthenelse{\equal{#1}{E}}{Functions}{%
  \ifthenelse{\equal{#1}{B}}{Sets}{}}}}}%
  ]}
\usepackage{microtype}

\begin{document}
	


\title{\bf\LARGE Communication-aware Wide-Area Damping Control \\ using Risk-Constrained Reinforcement Learning}
\author{
\IEEEauthorblockN{Kyung-bin Kwon, Lintao Ye, Vijay Gupta, and Hao Zhu}

\thanks{\protect\rule{0pt}{3mm} This work has been partially supported by NSF Grants 2130706, 2150571, 2222097, and 2300355, and by ARO Grant W911NF2310266.}
\thanks{\protect\rule{0pt}{3mm}  K.~Kwon is with the Optimization and Control Group, Pacific Northwest National Laboratory, Richland, WA 99352, USA; Email: kyung-bin.kwon@pnnl.gov.

L.~Ye is with the School of Artificial Intelligence and Automation, Huazhong University of Science and Technology, Wuhan 430074, China; Email: yelintao93@hust.edu.cn.

V.~Gupta is with the Elmore School of Electrical and Computer Engineering, Purdue University, West Lafayette, IN 47906, USA; Email: gupta869@purdue.edu.

H.~Zhu is with the Chandra Family Department of Electrical \& Computer Engineering, The University of Texas at Austin, Austin, TX 78712, USA; Email: haozhu@utexas.edu.
}}

\markboth{IEEE Transactions on Smart Grid (Accepted)}%
{Kwon \MakeLowercase{\textit{et al.}}: Communication-aware Wide-Area Damping Control using Risk-Constrained Reinforcement Learning}
\renewcommand{\thepage}{}
\maketitle
\pagenumbering{arabic}

\begin{abstract}

Non-ideal communication links, especially delays, critically affect fast networked controls in power systems, such as the wide-area damping control (WADC). Traditionally, a delay estimation and compensation approach is adopted to address this cyber-physical coupling, but it demands very high accuracy for the fast WADC and cannot handle other cyber concerns like link failures or {cyber perturbations}. Hence,  we propose a new risk-constrained framework that can target the communication delays, yet amenable to general uncertainty under the cyber-physical couplings. Our WADC model includes the synchronous generators (SGs), and also voltage source converters (VSCs) for additional damping capabilities. 
To mitigate uncertainty, a mean-variance risk constraint is introduced to the classical optimal control cost of the linear quadratic regulator (LQR). Unlike estimating delays, our approach can effectively mitigate large communication delays by improving the worst-case performance. 
A reinforcement learning (RL)-based algorithm, namely, stochastic gradient-descent with max-oracle (SGDmax), is developed to solve the risk-constrained problem. We further show its guaranteed convergence to stationarity at a high probability, even using the simple zero-order policy gradient (ZOPG). Numerical tests on the IEEE 68-bus system not only verify  SGDmax's convergence and VSCs' damping capabilities, but also demonstrate that our approach outperforms conventional delay compensator-based methods under estimation error. While focusing on performance improvement under large delays, our proposed risk-constrained design can effectively mitigate the worst-case oscillations, making it equally effective for addressing other communication issues and cyber perturbations. 
\end{abstract}


\section{Introduction}\label{sec:IN}
Wide-area damping control (WADC) is increasingly critical for mitigating inter-area oscillations arising from renewable penetration and large power transfers 
\cite{pnnl}. Despite recent advancements of wide-area measurement system (WAMS), the WADC implementations remain challenged due to  new energy resources \cite{shi, kumar, ps_res1, ps_res2}, and more importantly, its reliance on  fast communication links for issues like  communication delays \cite{bento, zhou, ps_delay1, ps_delay2} and {cyber perturbations} \cite{mahapatra, patel, zhao, ps_attack1}.

The optimal WADC problem can be formulated using the linear quadratic regulator (LQR) objective, which minimizes the deviations of state variables such as frequency and angle in addition to the control effort \cite{pandey}. The underlying communication network of WAMS allows the controller to map from the state data provided by phasor measurement units (PMUs) to individual actuation points. Thus, the impact of communication technologies on WADC is mainly two-fold. First, in practice, the WAMS communication links are limited in numbers, and the network connectivity either follows a {\em structured feedback} \cite{dorfler} or an aggregated scheme \cite{sayak}. While we consider the former by following a prescribed connectivity between the actuation inputs and the PMU data, our approach is also generalizable to the latter. Second, and more importantly,  the fast control timescales and high sampling rates introduce link-specific delays and  asynchronous inputs to the distributed actuators \cite{stahlhut}. 
These random yet heterogeneous delays, like other cyber concerns, can increase system variability especially at the price of worst-case performance.

To mitigate communication delays, recent WADC research mainly  relies on the adaptive delay compensators (ADCs). Specifically, in \cite{shi} and \cite{darabian}, ADCs are employed to offset the WAMS delays using techniques such as one-step-ahead predictive control or $H_\infty/H_2$ synthesis. Additionally, \cite{bento} and \cite{alghamdi} analyze the impact of maximum and minimum time delays on system dynamics, while \cite{liu_pd,nayak_pd} design ADCs by accounting for the stochastic nature of communication delays. However, the effectiveness of all these approaches critically depends on the accuracy of delay estimation. Typically, delays are estimated using time tags in data packets \cite{padhy}, but the accuracy can be easily compromised by external factors like network disruptions, further worsened by potential cyber intrusions \cite{patel_ca, zhao_ca}. Moreover, the estimation errors are known to accumulate across time \cite{shi}, and WADC is extremely sensitive to this issue at the price of prolonged control horizons. All of these factors motivate us to develop a new paradigm of communication-aware WADC that does not rely on precise delay estimation. \textcolor{black}{Table~\ref{tb:comparison} has listed a detailed comparison between WADC-related literature and our proposed risk-constrained approach,  highlighting the strength of our work in bypassing the need of precise delay information.} More excitingly, our approach can tackle general perturbations regardless of the exact sources, and thus can mitigate other cyber issues of great concern to WADC, including link failures \cite{ps_delay1, ps_delay2} or cyber intrusions \cite{mahapatra, mcgill}. 

\begin{table*}[ht]
\centering
\resizebox{\textwidth}{!}{%
\begin{tabular}{@{}lllll@{}}
\toprule
\textbf{Controller}         & \textbf{Communication Delay Handling}     & \textbf{Communication Failure Handling}      & \textbf{Model Dependency}       & \textbf{Delay Information} \\ \midrule
\textbf{\cite{shi}}         & Adaptive delay compensator                & Not addressed                                & Model-free                      & Needed                                   \\ 
\textbf{\cite{kumar}}       & Static delays considered                  & Not addressed                                & Model-based                     & Needed                                   \\ 
\textbf{\cite{ps_delay1}}   & Robust H-$\infty$                         & Limited robustness                           & Model-based                     & Needed                                   \\ 
\textbf{\cite{ps_delay2}}   & Variable delays considered                & Packet dropout considered                    & Model-based                     & Needed                                   \\ 
\textbf{\cite{mahapatra}}   & Not addressed                             & 
{Cyber perturbations} addressed                        & Model-based                     & Not applicable                           \\ 
\textbf{\cite{dorfler}}     & Not addressed                             & Not addressed                                & Model-based                     & Needed                                   \\ 
\textbf{\cite{sayak}}       & Not addressed                             & Partially addressed                          & Model-free                      & Not applicable                           \\ 
\textbf{\cite{stahlhut}}    & Latency explicitly modeled                & Not addressed                                & Model-based                     & Needed                                   \\ 
\textbf{\cite{alghamdi}}    & Variable delays considered                & Not addressed                                & Model-based                     & Needed                                   \\ 
\textbf{\cite{liu_pd}}      & Adaptive delay compensator                & Not addressed                                & Model-free                      & Needed                                   \\ 
\textbf{\cite{nayak_pd}}    & Adaptive-probabilistic delay compensator  & Limited robustness                           & Model-free                      & Needed                                   \\ 
\textbf{\cite{mcgill}}      & Not addressed                             & {Cyber perturbations} addressed                        & Model-free                      & Not applicable                           \\ 
\textbf{\cite{beiraghi}}    & Adaptive delay compensator                & Not addressed                                & Model-based                     & Needed                                   \\ 
\textbf{\cite{cheng}}       & Adaptive delay compensator                & Not addressed                                & Model-based                     & Needed                                   \\ 
\textbf{Proposed Method}    & Risk-constrained control                        & Robustness addressed                           & Model-based                     & \textbf{Not needed}                              \\ \bottomrule
\end{tabular}%
}
\caption{Comparison of Proposed Risk-constrained Design with Existing WADC Approaches}
\label{tb:comparison}
\end{table*}

%

There are two recent trends in WADC designs beyond mitigating delays. First, 
reinforcement learning (RL) techniques, both model-based or model-free, are popularly adopted, including Q-learning \cite{duan}, deep neural networks \cite{gupta}, and actor-critic methods \cite{hashmy}.
While model-free RL does not need the system knowledge, it demands extensive data and exhibits slow learning rates for large-scale problems \cite{tu,lintao}. In contrast, the model-based RL leverages either known or estimated system models \cite{sayak,mcgill,thakallapelli} to generate offline simulations to attain fast convergence and safe controls. This trade-off motivates us to adopt model-based RL  for the WADC problem as fast training and online safety are very important therein.
Second, integrating voltage source converters (VSCs) enhances WADC's damping support \cite{mpce_vsc}. While prior work \cite{mcgill, pang} has modeled flexible VSC power outputs, synchronous generator (SG) models omit the crucial exciter component, which requires a new approach for the system modeling.
%

This paper designs a risk-constrained RL-based WADC approach that explicitly targets the issue of communication delays but can generalize other cyber perturbations. 
We develop a new linearized system dynamic model with both VSCs and fourth-order SG dynamics, since it is necessary to include the exciters and a small-signal analysis is sufficient for WADC. This way, an LQR-based optimal control problem with structured feedback is formulated per the communication network's connectivity. 
Our analysis suggests the level of perturbation to system dynamics increases with the delay, and so does the variability of LQR cost. This inspired us to put forth a mean-variance risk constraint to bound the variability of state cost. To solve the resultant constrained problem, we develop a stochastic gradient-descent with max-oracle (SGDmax) to approach the stationary point of the dual problem. This model-based RL method will use the zero-order policy gradient (ZOPG) to simplify the gradient estimation, which is still guaranteed to converge at a high probability. Numerical tests have shown our approach can outperform conventional delay compensators-based methods, as the latter critically depends on the accuracy of delay estimation and other perturbations. 
\textcolor{black}{While our risk-constrained WADC design focuses on mitigating the impact of communication delays, it can be used to deal with other imperfect communication scenarios like packet drops, losses, and disorder \cite{ps_delay1, padhy2}. This is because imperfect communications can be similarly modeled as an additional disturbance to the system input signals, which can be effectively mitigated by our approach via reducing the variance risk.} {\KBD This  generalizability has been demonstrated in our related work on risk-aware grid-forming inverter (GFM) control \cite{kwon2023risk} for mitigating large load disturbances. However, there exist significant differences in the modeling of \cite{kwon2023risk}, and it remains open to develop a new risk-constrained WADC formulation  to address the unique challenges posed by time-varying delays.}

To sum up, the main contributions of our work are two-fold. First, we have
introduced a new paradigm of communication-aware WADC by systematically addressing the impacts of non-negligible delays {\KBC without the need to estimate delays}. Our design can effectively increase the worst-case damping level even under large delays or general cyber perturbations. \textcolor{black}{Second, we formulate the problem of optimizing the structured feedback which enables the application of the SGDmax algorithm, capitalizing on its sample efficiency and low online computations over model-free RL,  and guaranteed convergence to attain robust, risk-constrained WADC performance.}
In short, our risk-constrained WADC design can verifiably enhance the performance under large delays, thereby providing a new direction of pursuing communication-aware controls with the increasing cyber-physical coupling of next-generation power systems.

The rest of this paper is structured as follows. Section~\ref{sec:MW} formulates the linearized system model that includes VSCs. In Section~\ref{sec:risk}, we model the communication networks and analyze the impacts of delays, as well as formulate the risk-constrained LQR problem. Section~\ref{sec:SGD} presents the RL-based SGDmax algorithm and its convergence analysis. In Section~\ref{sec:NT}, numerical tests on the IEEE 68-bus system will be used to demonstrate the performance improvements of the proposed design, and the paper is concluded in Section~\ref{sec:CN}.


\section{System Modeling}\label{sec:MW}
\begin{figure}[t]
	\centering
	\includegraphics[width=\linewidth]{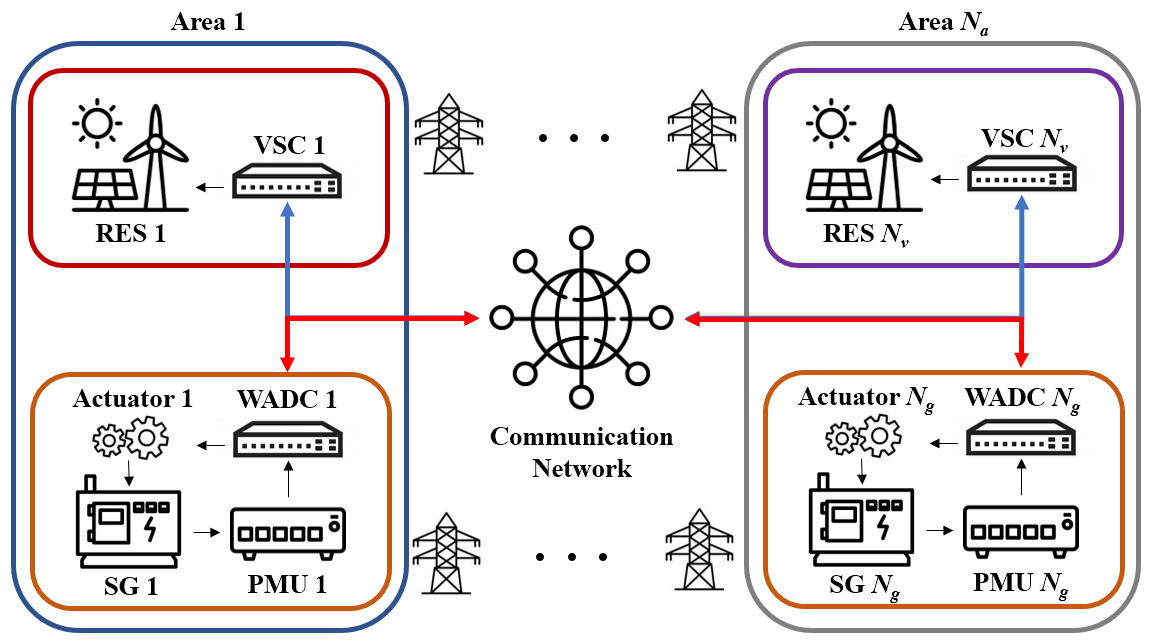}
	\caption{System depiction with synchronous generators (SGs) and voltage source converters (VSCs) participating in WADC.} 
	\label{fig:system}
\end{figure}

We consider a power system partitioned into $N_a$ areas as shown in Fig.~\ref{fig:system}. It consists of $N_g$ synchronous generators (SGs), as indexed by the set $\mathcal{N} = \{1,2,\dots,N_g\}$. We assume that all SGs are equipped with phasor measurements units (PMUs) and wide-area damping controllers (WADC). Only a subset of SGs participating in WADC is also possible, by eliminating certain SGs from the feedback design.  In addition, the state of SG $i$ can be measured by its local PMU, denoted by $\mathbf x_{i} = [\delta_i, \omega_i, E_i, E^{fd}_{i}]^\intercal$. 
The electro-mechanical (EM) states $\delta_i$ and $\omega_i$ denote the deviation of internal rotor angle and speed from the operating point, respectively; while the non-EM states, $E_i$ and $E^{fd}_i$, indicate the generator internal voltage and excitation voltage, respectively  \cite{sayak}. While our analysis focuses on the fourth-order model,  the numerical tests in Section \ref{sec:NT} will consider the more practical sixth-order SG dynamics.

The fourth-order dynamics for SG $i\in\mathcal{N}$  is represented by: 
\begin{subequations}
\begin{align}
    &\dot{\delta}_i = \omega_i,\label{eq:di}\\
    &\dot{\omega}_i = \frac{1}{2H_i}\left[P^m_{i}-P^e_{i}-D_i \omega_i \right],\label{eq:wi}\\
    &\dot{E}_i \!=\! \frac{1}{T_i^{d'}}\!\left[-\frac{x^d_{i}}{x_i^{d'}}E_i\! +\! (x^d_{i}-x_i^{d'}) I^d_i+E^{fd}_{i} \right], \label{eq:ei}\\
    &\dot{E}^{fd}_{i}=\frac{1}{T^a_{i}}[-E^{fd}_{i}-K^a_{i}(E_i - x_i^{d'} I^d_i - \bar{V}_{i}-\Delta\bar{V}_{i})]. \label{eq:efd}
\end{align} \label{eq:gen}
\end{subequations}
\noindent The mechanical power input $P^m_i$ and reference voltage $\bar{V}_i$ are considered fixed as they are determined by slower operations than WADC. The electric power output $P^e_i$ and d-axis current $I^d_i$ are algebraic variables depending on the full network nonlinear power flow, which will be linearized later on. Notably, the damping control signal $\Delta \bar{V}_i$ in \eqref{eq:efd} can quickly adjust the excitation voltage ${E}^{fd}_{i}$ and affect other SG states, in order to improve the damping performance. The other terms like $H_i$ and $D_i$ are constant parameters given by generator specifications, with the reference voltage $\bar{V}_i$ determined by longer-timescale operations such as optimal power flow. 


Along with SGs, the system also has $N_v$ voltage source converters (VSCs) indexed by the set $\mathcal{V} = \{N_g+1, N_g+2, \ldots, N_g+N_v\}$ that can participate in WADC. \textcolor{black}{The VSC-connected energy sources are modeled as power sinks and sources that generate active and reactive power to the grid. Here, VSCs can quickly provide supplementary active and reactive power adjustments around their steady-state references \cite{fan}.} For each VSC $j \in \mathcal{V}$, we control its power injection adjustments $\Delta {P}_j^v$ and $\Delta {Q}_j^v$.

To formulate the overall network dynamics, we need to consider the power flow coupling between the SG internal nodes and VSC buses. This allows us to express $P^e_i$ and $I^d_i$ in \eqref{eq:gen} as functions of the full system states and control inputs. 
By applying Kron reduction \cite{kron} and eliminating all other buses,  we consider the power flow equations between all SG interval voltages $\{E_i, \delta_i\}_{i\in\mathcal{N}}$ and  all VSC terminal bus voltages $\{V_j, \theta_j\}_{j\in\mathcal{V}}$. 
For each SG $i\in \mathcal{N}$, the active and reactive power outputs are given by
\begin{align*}
   P^e_{i} &= \textstyle \sum_{\ell=1}^{N_g} E_i E_\ell (G_{
i\ell} \cos(\delta_i - \delta_\ell) + B_{i\ell} \sin(\delta_i - \delta_\ell)) \nonumber\\ &+ \textstyle \sum_{\ell=N_g+1}^{N_g+N_v} E_i V_\ell (G_{i\ell} \cos(\delta_i - \theta_\ell) + B_{i\ell} \sin(\delta_i - \theta_\ell)), \\
    Q^e_{i} &= \textstyle \sum_{\ell=1}^{N_g} E_i E_\ell (G_{i\ell} \sin(\delta_i - \delta_\ell) - B_{i\ell} \cos(\delta_i - \delta_\ell)) \nonumber\\ &+ \textstyle \sum_{\ell=N_g+1}^{N_g+N_v} E_i V_\ell (G_{i\ell} \sin(\delta_i - \theta_\ell) - B_{i\ell} \cos(\delta_i - \theta_\ell)). 
\end{align*} 
\noindent Similar equations can be written for each VSC node $j\in\mathcal V$, namely $P^v_j$ and $Q^v_j$.
To match with the SG dynamics in \eqref{eq:gen}, we should also consider the d-axis current flow  $I^d_i = Q^e_i/E_i$ using the reactive power output, namely
\begin{align*}
	I^d_{i} &= \textstyle \sum_{\ell=1}^{N_g} E_\ell (G_{i\ell} \sin(\delta_i - \delta_\ell) - B_{i\ell} \cos(\delta_i - \delta_\ell)) \nonumber\\ &+ \textstyle \sum_{\ell=N_g+1}^{N_g+N_v} V_\ell (G_{i\ell} \sin(\delta_i - \theta_\ell) - B_{i\ell} \cos(\delta_i - \theta_\ell)). 
\end{align*}

{\KBC We consider a linearized dynamic model which would facilitate the analysis of risk constraint later on, based on a fixed steady-state operating point.} A linearized model could well capture the WADC dynamics~\cite{mcgill}, and we will test the proposed design on the actual nonlinear system model in numerical simulations. By using the bold symbols to concatenate all variables into the vector form, we have
\renewcommand\arraystretch{1.3}
\begin{align}
    \begin{bmatrix} \Delta \mathbf{P}^e \\ \Delta \mathbf{I}^d \\ \Delta \mathbf{P}^v \\ \Delta \mathbf{Q}^v \end{bmatrix} = \begin{bmatrix} 
    	\frac{\partial \mathbf{P}^e}{\partial \bm{\delta}} & \frac{\partial \mathbf{P}^e}{\partial \mathbf{E}} & \frac{\partial \mathbf{P}^e}{\partial \bm{\theta}} & \frac{\partial \mathbf{P}^e}{\partial \mathbf{V}}  
    	\\ \frac{\partial \mathbf{I}^q}{\partial \bm{\delta}} & \frac{\partial \mathbf{I}^q}{\partial \mathbf{E}} & \frac{\partial \mathbf{I}^q}{\partial \bm{\theta}} & \frac{\partial \mathbf{I}^q}{\partial \mathbf{V}} 
    	\\ \frac{\partial \mathbf{P}^v}{\partial \bm{\delta}} & \frac{\partial \mathbf{P}^v}{\partial \mathbf{E}} & \frac{\partial \mathbf{P}^v}{\partial \bm{\theta}}  & \frac{\partial \mathbf{P}^v}{\partial \mathbf{V}} 
    	\\ \frac{\partial \mathbf{Q}^v}{\partial \bm{\delta}} & \frac{\partial \mathbf{Q}^v}{\partial \mathbf{E}} & \frac{\partial \mathbf{Q}^v}{\partial \bm{\theta}}  & \frac{\partial \mathbf{Q}^v}{\partial \mathbf{V}} \end{bmatrix} 
     \begin{bmatrix} \Delta\bm{\delta}\\ \Delta \mathbf{E} \\ \Delta\bm{\theta}\\ \Delta \mathbf{V} \end{bmatrix},  \label{eq:delta}
\end{align}
\renewcommand\arraystretch{1.0}

\noindent where the partial derivatives form the Jacobian matrix. This way, we can represent the algebraic variables $\Delta \mathbf{P}^e$ and $\Delta \mathbf{I}^d$ as follows; see the derivations in \cite[Appendix]{kwon_wadc}.
\begin{subequations}
\begin{align}
	&\Delta \mathbf{P}^e = \mathbf{A}^P_1 \Delta\bm{\delta} + \mathbf{A}^P_2 \Delta \mathbf{E} + \mathbf{A}^P_3 \Delta \mathbf{P}^v + \mathbf{A}^P_4 \Delta\mathbf{Q}^v,  \label{eq:pe} \\
	&\Delta \mathbf{I}^d =\mathbf{A}^I_1 \Delta\bm{\delta} + \mathbf{A}^I_2 \Delta \mathbf{E} + \mathbf{A}^I_3 \Delta \mathbf{P}^v + \mathbf{A}^I_4 \Delta\mathbf{Q}^v.  \label{eq:id}
\end{align} \label{eq:peid}
\end{subequations}
The linearized relation in \eqref{eq:peid} will allow to integrate all SGs' dynamics in  \eqref{eq:gen}  with the VSC power injections. For simplicity, our work does not consider the VSC's reactive power adjustment by fixing $\Delta \mathbf Q^v = 0$. This is because the reactive power component has much smaller impact on WADC than the active one, as discussed in \cite{q1,q2}. 

By substituting \eqref{eq:peid} into \eqref{eq:gen}, the overall dynamics for $\mathbf{x} := \{\mathbf{x}_i\}_{i\in\mathcal{N}}$ with the full input $\mathbf{u} := [(\Delta \bar{\mathbf{V}})^\intercal ~ (\Delta \mathbf{P}^v)^\intercal]^\intercal \in \mathbb{R}^{N_g + N_v}$ can be linearized as: 
\begin{align}
    \dot{\mathbf{x}} = \mathbf{A}_c \mathbf{x} + \mathbf{B}_c  \mathbf{u} + \bm{\xi}\;
   \label{eq:vsc_dy}
\end{align}
\textcolor{black}{where $\bm{\xi} \in \mathbb{R}^{4N_g}$ captures the overall random perturbation to the dynamic model. Our analysis later only requires to know the moments  of $\bm{\xi}$, not its specific distribution, but this perturbation could arise from Gaussian noise from sensing, imperfect communications, and external disturbances.} Note that SG states in $\mathbf{x}$ are actually deviations from the corresponding steady-state values due to the linearization.  
Last, the rest of the paper will consider the discrete-time dynamics based on \eqref{eq:vsc_dy}, given by
\begin{align}
    \mathbf{x}_{t+1} = \mathbf{A} \mathbf{x}_t + \mathbf{B} \mathbf{u}_t + \bm{\xi}_t,~\forall t =0,1,\ldots \label{eq:dyn}
\end{align}
which has been obtained with a sufficiently small time step. \textcolor{black}{Note that the proposed method is robust to the changes in the operating point 
as the system matrices $\mathbf{A}$ and $\mathbf{B}$ can be updated accordingly. Furthermore, operating points in power systems typically change gradually and within a predictable range, ensuring the validity of the linearized model for most practical scenarios. A numerical test involving perturbations of the operating point has been presented in Section~\ref{sec:NT} to demonstrate the effectiveness of our proposed method under varying operating conditions.}


\begin{remark}[WADC dynamics with controllable VSCs] The key to establishing the overall system model \eqref{eq:vsc_dy} lies in integrating the VSC power injections with the SG dynamics in \eqref{eq:gen}.  
Different from earlier work \cite{mcgill}  using the second-order swing equations for SG dynamics, our work  instead employs the fourth-order SG dynamics  which are more accurate and better connected to excitation control \cite{gmodel}. Nonetheless, the latter makes it more challenging to integrate the VSCs which can affect both $\mathbf{P}^e$ and $\mathbf{I}^d$ at the SG terminals.  To this end, we have considered the static power flow coupling between the SG internal voltages and the VSC voltages, and used the linearization approach to simplify the coupling. Thus, we have improved the modeling accuracy for representing the VSC-integrated system dynamics and designing the WADC. 


\end{remark}


\section{Risk-constrained WADC Problem} \label{sec:risk}
For the system \eqref{eq:dyn}, we can formulate the WADC problem as a linear quadratic regulator (LQR), to solve
\begin{align}
 R_0(\mathbf{K}) &\! = \!\lim_{T \to \infty}\frac{1}{T}\mathbb{E} \sum_{t=0}^{T-1} [\mathbf{x}^\intercal_t \mathbf{Q} \mathbf{x}_t + \mathbf{u}^\intercal_t \mathbf{R} \mathbf{u}_t]  \label{eq:obj}
\end{align}
\textcolor{black}{where the expectation is taken with respect to the random perturbation $\bm{\xi}_t$. Both  $\{\mathbf{Q}, \mathbf{R}\}$ are given positive (semi-)definite matrices to weigh the corresponding elements, as similarly done in earlier WADC work \cite{mcgill, alghamdi}.} \textcolor{black}{Note that the $\mathbf{Q}$ and $\mathbf{R}$ matrices 
can be tailored to specific operational goals, allowing for adjustments such as emphasizing particular state variables or minimizing control effort based on system priorities.} The decision variable here is the feedback gain $\mathbf{K} \in \mathbb{R}^{(N_g+N_v) \times (4N_g)}$ for a linear mapping between ${\mathbf{x}}_t$ and $\mathbf{u}_t$; i.e., $\mathbf{u}_t = - \mathbf{K} \mathbf{x}_t$. 
%
A structured feedback $\mathbf{K} \in \mathcal{K}$ is very common in WADC due to limited deployment of communication links. Thus, for a given sparse 
communication graph between any SG $i\in\mathcal{N}$ and control node $\ell\in\mathcal{N}\cup\mathcal{V}$, the feasible set $\mathcal{K}$ becomes
\begin{align}
    \mathcal{K} = \{\mathbf{K}: \mathbf{K}_{\ell,i} = 0 \;\text{if and only if}\; \ell \nleftrightarrow i) \} \nonumber
\end{align}
where $\ell \nleftrightarrow i$ indicates no communication link available from SG $i$ to control node $\ell$. Note that  
$\mathbf{K}_{\ell,i} \in \mathbb{R}^{1 \times 4}$ denotes the submatrix of $\mathbf{K}$ mapping from $\mathbf{x}_i$ to $\mathbf{u}_\ell$. 
While the sparsity constraint on $\mathbf{K}$ makes it difficult to analyze the feasible region \cite{feng}, 
it will not affect the implementation of our proposed gradient-based solutions as detailed later on.  

Furthermore, the communication delays through the WAMS are a crucial factor affecting the performance of WADC \cite{zhou,stahlhut}. Due to the very fast timescale of WADC, typically at 0.01s level, the communication delay effects are more notable than other slower control designs. 
To model it, we consider that the measured $\mathbf{x}_{i,t}$ at SG $i$ would experience a time-invariant delay $h_i$ when reaching all other control nodes $\ell \neq i$.  Per time $t$, let us denote this delayed state by $\tilde{\mathbf{x}}_{i,t} := \mathbf{x}_{i,t-h_{i}}$. This way, the local state vector available at each control node $\ell \in \mathcal{N} \cup \mathcal{V}$ becomes $\tilde{\mathbf{x}}^{(\ell)}_{t} = [\tilde{\mathbf{x}}_{1,t}^\intercal, \ldots, \mathbf{x}_{\ell,t}^\intercal, \ldots, \tilde{\mathbf{x}}_{N_g,t}^\intercal]^\intercal$ which uses all delayed states except for the local SG state if $\ell \in \mathcal{N}$. 
For simplicity, our model assumes a uniform communication delay for each SG's state with the fact that WADC relies on optical fiber cables for communications \cite{padhy2}. However, we can generalize it to the heterogeneous delay setting with different (or even random)  delay times for each link and utilize our risk-constrained WADC design to address more realistic settings.  


The communication delays are detrimental to maintaining the WADC performance, as they introduce additional uncertainty and perturbations to the system dynamics in \eqref{eq:dyn}. Intuitively, with larger delays, the delayed state $\tilde{\mathbf{x}}^{(\ell)}$ received at control node $\ell$ would incur a higher error difference from the actual state. As a result, the control input $\mathbf{u}_\ell$ formed by $\tilde{\mathbf{x}}^{(\ell)}$ would introduce an increasing perturbation to the system dynamics. Specifically,  for the control $\mathbf{u}_{\ell,t}$ per time $t$, using the delayed state $\tilde{\mathbf{x}}^{(\ell)}_t$  would cause the perturbation as follows:
\begin{align}
    \beta_{\ell,t} = \sum_{i\in \mathcal{N}, i \neq \ell} \mathbf{K}_{\ell,i} (\tilde{\mathbf{x}}_{i,t}- \mathbf{x}_{i,t} ).\label{eq:beta}
\end{align}
Note that the difference term for each SG's state  $(\tilde{\mathbf{x}}_{i,t}- \mathbf{x}_{i,t})$ would accumulate as the delay timing $h_i$ increases. 
\textcolor{black}{To incorporate  $\beta_{\ell,t}$, one can update the perturbation term in \eqref{eq:dyn} to $\bm\xi'_t$ which aggregates the effects of both the original  $\bm{\xi}_t$ and  this delay-induced perturbation. With large delays,  non-negligible $\beta_{\ell,t}$ significantly increases the level of variability  for \eqref{eq:dyn}, challenging the LQR-based WADC design which only focuses on average trajectory performance. }

To address this delay-induced perturbation, we propose a risk-constrained LQR  formulation for the WADC problem by limiting the mean-variance risk of the state deviation, as
\begin{align}
	&\min_{\mathbf{K} \in \mathcal{K}} \; R_0(\mathbf{K}) \! = \!\lim_{T \to \infty}\frac{1}{T}\mathbb{E} \sum_{t=0}^{T-1} [\mathbf{x}^\intercal_t \mathbf{Q} \mathbf{x}_t + \mathbf{u}^\intercal_t \mathbf{R} \mathbf{u}_t] \label{eq:opt}\\
	&\textrm{s.t.}~R_c(\mathbf{K}) \! =\!\! \lim_{T \to \infty}\!\frac{1}{T} \mathbb{E} \sum_{t=0}^{T-1} \big(\mathbf{x}_t^\intercal \mathbf{Q} \mathbf{x}_t - \mathbb{E}[\mathbf{x}_t^\intercal \mathbf{Q} \mathbf{x}_t \vert \mathcal{F}_t]\big)^2 \leq c, \nonumber
\end{align}
%
where $\mathcal{F}_t := \{\mathbf{x}_0, \mathbf{u}_0, \ldots, \mathbf{x}_{t-1}, \mathbf{u}_{t-1}\}$ collects 
the system trajectory up to time $t$, while the scalar $c$ is a risk tolerance threshold that can be set based on the allowable level of state fluctuation. 

A smaller $c$ enforces stricter constraints, reducing state variability but increasing control effort, while a larger $c$ provides more flexibility at the cost of stability. This mean-variance risk constraint aims to reduce the average deviation of the state cost term $(\mathbf{x}_t^\intercal \mathbf{Q} \mathbf{x}_t)$ from its expected value conditioned on the past data, thus mitigating the high system variability due to external perturbations.
\textcolor{black}{It is noteworthy that by incorporating the constraint, we can account for the effects of communication delay variability by limiting the impact of delay-induced state fluctuations within a predefined risk tolerance threshold.} 
{\KBD Note that a similar risk-constrained approach has been successfully applied to the control of GFMs to tackle uncertainty in load variation \cite{kwon2023risk}. Since the communication delays introduce a unique perturbation pattern as shown in \eqref{eq:beta}, both applications highlight the versatility of the proposed risk constraint in handling various sources of uncertainty in fast power system controls.}
The benefit of using it for the WADC problem is two-fold, as discussed in the following remark.

\begin{remark}[Risk-constrained WADC]
The mean-variance risk constraint can address the increased system perturbation and thus state variability due to the large communication delays in fast WADC problems. 
\textcolor{black}{It effectively limits the variability of 
the state-related cost term from its expected value conditioned on past data, thereby reducing its \textit{worst-case} cost which may arise due to variability in communication delays.}
This is very important for the resultant WADC designs to meet the safety operations limits in power system dynamics. {It is worth mentioning that this risk-constrained framework is much more preferred over a robust LQR design \cite{SCAMPICCHIO2021109571} which tends to be very conservative when considering all possible perturbations in the uncertainty set; see e.g., \cite{tsiamis,tsiamis2}.} \textcolor{black}{Moreover, by focusing on reducing statistical dispersion rather than guarding against all extremes, the mean‑variance approach maintains strong nominal damping performance and mitigates risk under uncertainty, avoiding the excessive conservativeness of robust LQR designs.}
\end{remark}


While the risk term $R_c(\mathbf{K})$  appears more complicated than the LQR cost, it can be expressed in a quadratic form as analyzed in \cite{tsiamis,tsiamis2}. {\KBC Under the linearized dynamics and assuming  a finite fourth-order moment for $\bm{\xi}'_t$, we have:}
\begin{align}
	R_c(\mathbf{K}) \!= \!\! \lim_{T \rightarrow \infty} \frac{1}{T} \mathbb{E} \sum_{t=0}^{T-1} \Big(4\mathbf{x}_t^\intercal \mathbf{Q}\mathbf{W}\mathbf{Q} \mathbf{x}_t+4\mathbf{x}_t^\intercal \mathbf{Q}\mathbf{M}_3 \Big) \!\leq\! \bar{c} \label{eq:rconst2}
\end{align}
by defining $\bar{c} := c-m_4 + 4 \text{tr}\{(\mathbf{W}\mathbf{Q})^2 \}$ and 
\begin{align}
	\bar{\bm{\xi}'}&=\mathbb{E}[\bm{\xi}'_t], \mathbf{W}=\mathbb{E}[(\bm{\xi}'_t-\bar{\bm{\xi}}')(\bm{\xi}'_t-\bar{\bm{\xi}}')^\intercal], \nonumber\\ 
	\mathbf{M}_3 &= \mathbb{E}[(\bm{\xi}'_t-\bar{\bm{\xi}}')(\bm{\xi}'_t-\bar{\bm{\xi}}')^\intercal \mathbf{Q} (\bm{\xi}'_t-\bar{\bm{\xi}}')], \nonumber\\
	m_4&=\mathbb{E}[(\bm{\xi}'_t-\bar{\bm{\xi}}')^\intercal \mathbf{Q} (\bm{\xi}'_t-\bar{\bm{\xi}}') - \text{tr}(\mathbf{W}\mathbf{Q})]^2. \nonumber
\end{align} 
Note that the hyperparameter $c$ can be pre-defined through numerical analysis, while the noise statistics $\bar{\bm\xi}'$, $\mathbf{W}$, $\mathbf{M}_3$, and $m_4$ are computed based on an offline analysis. The reformulated risk constraint in \eqref{eq:rconst2} will help the algorithmic development in the ensuing section.
\textcolor{black}{
\begin{remark}(Generalizing imperfect communications).
While our analysis focuses on the communication delays that result in \eqref{eq:beta}, the proposed approach can generalize to other communication issues by appropriately setting up the perturbation term $\bm\xi'_t$ in \eqref{eq:rconst2}. Essentially, the latter can represent any possible mismatch between the ground-truth state (under perfect communications) and the actually received state at the actuator. 
This allows us to incorporate imperfect communications due to \textit{variable} time delays, packet dropouts, temporary denial-of-service attacks, or other network-induced uncertainty.
Therefore, our approach only needs to adopt an appropriate choice of the perturbation level for the actual communication scenario, and the same risk constraint in \eqref{eq:rconst2} would automatically ensure robustness to the exact imperfect patterns.
Section~V-D will illustrate this generalized robustness against a packet-loss scenario.
\end{remark}
}


\section{Learning the Risk-constrained WADC}\label{sec:SGD}


Following the procedures in \cite{GDA,kwon,kwon2023risk}, we present a gradient-based method for solving the risk-constrained WADC problem. \textcolor{black}{The constrained optimization in \eqref{eq:opt} poses some unique challenges for conventional RL solvers. This is because of both the inclusion of the risk constraint that makes unconstrained RL methods unsuitable, and the structured requirement on $\mathbf{K}$ leading to a complex geometry with disconnected feasible regions.} To address the first issue, we use the quadratic form of $R_c(\mathbf{K})$ in \eqref{eq:rconst2} to form the Lagrangian function by introducing a nonnegative multiplier $\lambda \geq 0$, defined as
\begin{align}
	&\mathcal{L}(\mathbf{K}, \lambda) 
    :=R_0(\mathbf{K})+ \lambda  [R_c(\mathbf{K})-\bar{c}]\nonumber\\
	=&\lim_{T \rightarrow \infty} \frac{1}{T}\mathbb{E}\sum_{t=0}^{T-1} \left[\mathbf{x}^\intercal_t \mathbf{Q}_{\lambda} \mathbf{x}_t\!+\!\mathbf{u}^\intercal_t \mathbf{R} \mathbf{u}_t\!+\!4 \lambda\mathbf{x}_t^\intercal \mathbf{Q} \mathbf{M}_3\right]\!-\!\lambda\bar{c}, \label{eq:lag}
\end{align}
where the weighted matrix is given by $\mathbf{Q}_{\lambda} := \mathbf{Q}+4\lambda (\mathbf{Q}\mathbf{W}\mathbf{Q})$.
This Lagrangian function is structurally similar to the standard LQR objective in \eqref{eq:obj}, with an additional linear term in $\mathbf{x}_t$. 

To simplify the adoption of gradient-based methods, consider the dual problem of  \eqref{eq:opt} as
\begin{align} 
\max_{\lambda\in\mathcal{Y}} \mathcal{D}(\lambda) = \max_{\lambda\in\mathcal{Y}}\min_{\mathbf{K}\in\mathcal{K}} \mathcal{L}(\mathbf{K},\lambda), \label{eq:maximin}
\end{align}
with $\mathcal{D}(\lambda) := \min_{\mathbf{K}\in\mathcal{K}} \mathcal{L}(\mathbf{K},\lambda)$ and the bounds $\mathcal{Y} := [0,\Lambda]$, where $\Lambda$ is chosen sufficiently large so that \eqref{eq:opt} is feasible. We consider the  minimax couterpart of \eqref{eq:maximin}:
\begin{align}
	\min_{\mathbf{K}\in\mathcal{K}} \Phi(\mathbf{K}), \quad \text{with} \quad \Phi(\mathbf{K}) := \max_{\lambda\in\mathcal{Y}} \mathcal{L}(\mathbf{K},\lambda). \label{eq:minimax}
\end{align}
Thanks to the linearity of $\mathcal{L}(\mathbf{K},\lambda)$ in $\lambda$, the inner maximization admits a closed-form solution (i.e., $\lambda=0$ if $R_c(\mathbf{K})\leq \bar{c}$ and $\lambda=\Lambda$ otherwise). Hence, solving  \eqref{eq:minimax} is more conveniently than the dual problem \eqref{eq:maximin}. Interestingly, under mild conditions, the KKT conditions for \eqref{eq:maximin} and \eqref{eq:minimax} are equivalent. This implies that the solution of \eqref{eq:minimax} will be a stationary point of \eqref{eq:maximin} \cite[Lemma~2]{kwon}, due to the nonconvexity of \eqref{eq:maximin}.

\subsection{Zero-Order Policy Gradient (ZOPG)}\label{subsec:ZOPG}

To optimize $\Phi(\mathbf{K})$ in \eqref{eq:minimax} via gradient updates, we need to (i) compute the gradient over the nonzero entries of $\mathbf{K}$ (denoted by $\nabla_{\mathcal{K}}$) to maintain the sparsity pattern and (ii) handle the non-differentiability arising from the discontinuous optimal $\lambda$ by using subgradients \cite{bu,kwon,GDA}.
However, direct computation of the gradient of $\Phi(\mathbf{K})$ is computationally expensive. Instead, we adopt the ZOPG method to estimate the gradient using finite differences, specifically, by evaluating the function with policy perturbation. 

To this end,  we can sample a perturbation $\mathbf{U}$ from $\mathcal{S}_{\mathcal{K}} = \{\mathbf{U}\in\mathcal{K}: \Vert \mathbf{U} \Vert = 1\}$, to guarantee that the perturbation retains to have the same sparsity structure as $\mathbf{K}$. Typically, the entries of $\mathbf{U}$ are drawn from a Gaussian or uniform distribution and then normalized. Using a smoothing radius $r>0$, the perturbed policy is $\mathbf{K}+r\mathbf{U}$, with  
\begin{align}
\Phi(\mathbf{K}+r\mathbf{U}) = \max_{\lambda\in\mathcal{Y}} ~\mathcal{L}(\mathbf{K}+r\mathbf{U},\lambda), \label{eq:perb}
\end{align}
where the best $\lambda$ value can be determined efficiently thanks to the linear dependency of $\mathcal{L}$ on $\lambda$. This finite difference provides an estimate of the gradient in the direction of $\mathbf{U}$. By scaling the estimate appropriately with the number of nonzero entries $n_\mathcal{K}$, we define the gradient estimator as
$\hat{\nabla}_{\mathcal{K}} \mathcal{L}(\mathbf{K};\mathbf{U}) = \frac{n_\mathcal{K}}{r}\, \mathcal{L}(\mathbf{K}+r\mathbf{U},\lambda')\, \mathbf{U} $,
with $\lambda'$ being the optimal $\lambda$ for \eqref{eq:perb}. This approach not only eliminates the need of explicit derivative computation but also naturally respects the structured sparsity of the controller.

%

\begin{algorithm}[t]
	\SetAlgoLined
	\caption{SGDmax with ZOPG for solving \eqref{eq:opt}}
	\label{alg:SGD}
	\DontPrintSemicolon
	{\bf Inputs:} The initial $\mathbf{K}^0$, upper bound $\Lambda$ for $\lambda$, step-size $\eta$,  the number of iterations $J$ and the number of ZOPG samples $M$.\; 
	\For{$j = 0, 1, \ldots, J-1$}{
		\For{$s=1,\ldots,M$}{
		Sample the random $\mathbf{U}_s \in \ccalS_\ccalK$. \;
        Implement ZOPG by obtaining $\lambda' \leftarrow \argmax_{\lambda \in \ccalY} \ccalL(\mathbf{K}^j+r\mathbf{U}_s, \lambda)$;\; 
	Estimate the gradient $\hat{\nabla}_{\ccalK} \ccalL (\mathbf{K}^j; \mathbf{U}_s) = \frac{n_\ccalK}{r} \ccalL(\mathbf{K}^j+r\mathbf{U}_s, \lambda') \mathbf{U}_s$.\;
		}
		Obtain the average $\hat{\mathbf{G}}(\mathbf{K}^j) = \frac{1}{M} \sum_{s=1}^M \hat{\nabla}_{\mathcal{K}}\ccalL(\mathbf{K}^j;\mathbf{U}_s)$. \;
		Update $\mathbf{K}^{j+1}\!\leftarrow\! \mathbf{K}^{j}\!-\!\eta \hat{\mathbf{G}}(\mathbf{K}^j)$.}
	{\bf Return:} the final iterate $\mathbf{K}^{J}$.
\end{algorithm}
\subsection{Stochastic Gradient-Descent with Max-Oracle (SGDmax)}\label{subsec:SGDmax}

\textcolor{black}{
To solve \eqref{eq:minimax}, we employ the SGDmax algorithm, which was first introduced in \cite{GDA}, leveraging ZOPG gradient estimates as outlined in Algorithm~\ref{alg:SGD}.} Starting from an initial feasible policy $\mathbf{K}^0$, SGDmax performs iterative updates of $\mathbf{K}$, where the \emph{max-oracle} refers to the inner maximization over $\lambda$ in \eqref{eq:minimax}.   To reduce the estimation variance of ZOPG, $M$ independent samples are used to find the averaged gradient  $\hat{\mathbf{G}}(\mathbf{K})$. The stochasticity of the ZOPG estimator  would make the  convergence analysis challenging. Thankfully, as shown in \cite[Appendix~B]{kwon}, if the iterates remain within a suitable sublevel set defined by the local Lipschitz and smoothness constants of $\Phi(\mathbf{K})$, then the algorithm converges to a stationary point with high probability.

\begin{proposition}\label{prop1}
By appropriately choosing the parameters $r$, $\eta$, \textcolor{black}{$M$}, and $J$ per the local Lipschitz and smoothness properties of $\Phi(\mathbf{K})$, and initializing with a feasible $\mathbf{K}^0 \in \mathcal{K}$, Algorithm~\ref{alg:SGD} converges to a stationary point of \eqref{eq:minimax} with high probability (approximately 90\%).
\end{proposition}


\section{Numerical Tests}\label{sec:NT}

\subsection{Simulation Setup}
\begin{figure}[t]
	\centering
	\includegraphics[width=\linewidth]{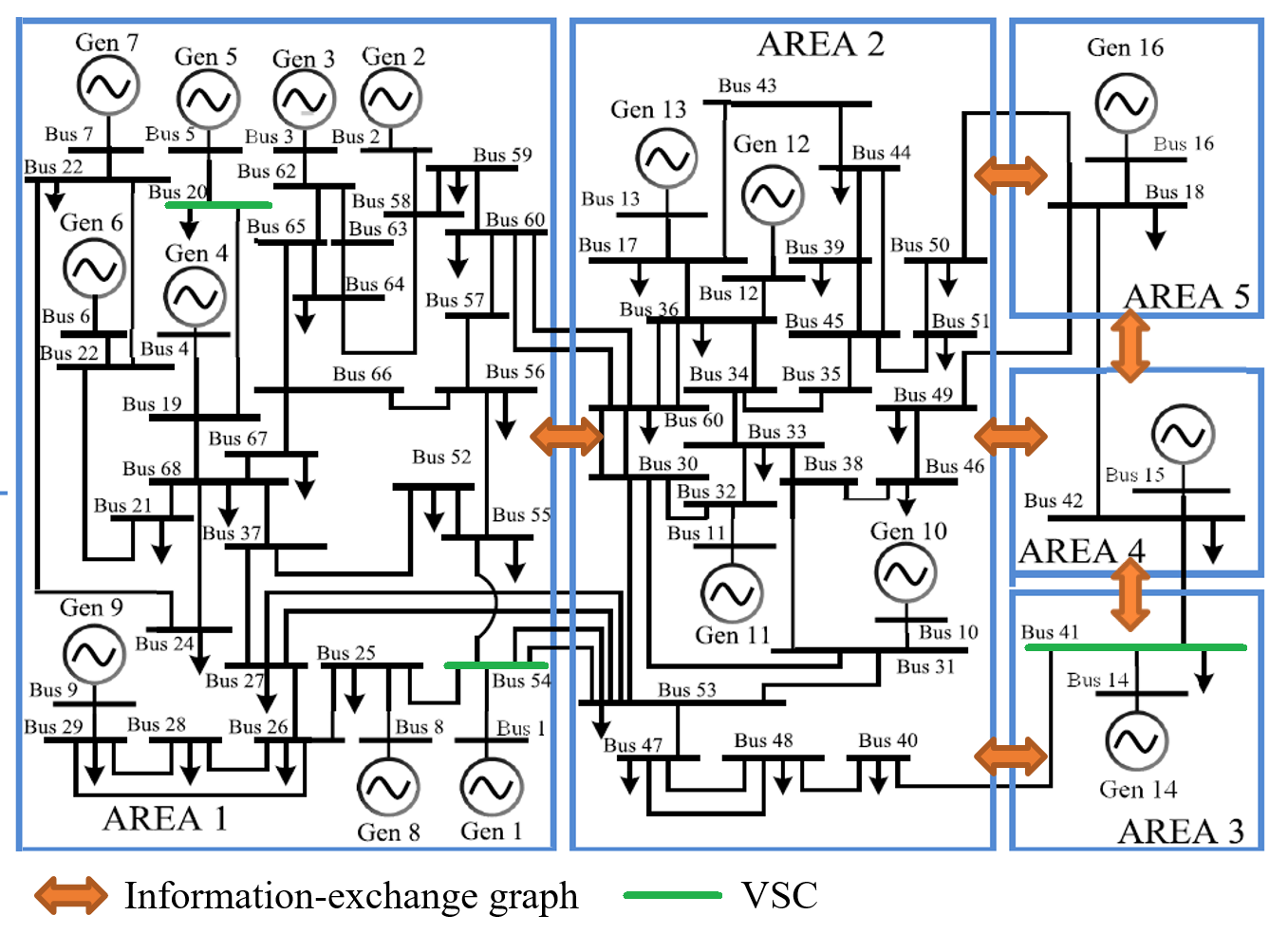}
	\caption{IEEE 68-bus test system with both SGs and VSCs.}
	\label{fig:IEEE68}
\end{figure}

\textcolor{black}{To demonstrate the effectiveness of our proposed risk-constrained WADC, we have conducted numerical tests on the IEEE 68-bus system using the parameters and load flow conditions given in \cite{ieee68}.} This system is a simplified model of the interconnection between the New York and New England power grids, consisting of five areas with a total of 16 SGs. Each SG is equipped with a WADC controller and a PMU meter. We also add three VSCs at buses 20, 42, and 54, similar to \cite{mcgill}. Based on the area partition in Fig.~\ref{fig:IEEE68}, the information is exchanged only between neighboring areas. For example, Area 2 can communicate with all other areas, while Area 1 can only exchange data with Area 2 but cannot access state measurements in Areas 3, 4, and 5. This makes us consider the structured feedback following the communication graph.

We have considered four types of WADC designs: \textit{SG}, \textit{SG-Risk}, \textit{VSC}, and \textit{VSC-Risk}, as listed in Table~\ref{tb:cases}. The \textit{SG} and \textit{SG-Risk} methods include SGs only for the WADC, while \textit{VSC} and \textit{VSC-Risk} have both SGs and VSCs participating in WADC. The risk-constraint has been considered in \textit{SG-Risk} and \textit{VSC-Risk}, with the other two developed using the unconstrained LQR cost. \textcolor{black}{Note that both \textit{SG} and \textit{VSC} are trained using Algorithm~\ref{alg:SGD} without any constraint multiplier $\lambda$, making both risk-neutral RL designs.}
\textcolor{black}{The training phase aims to compute the optimal WADC feedback gains under the given system dynamics. We first initialize the feedback gains and the risk-related parameter $\lambda$. Each training iteration is as follows:
\begin{enumerate}
    \item \textcolor{black}{Random impulse inputs are applied to all generators to model a sudden load step or fault, thereby creating a nonzero initial deviation state $\mathbf{x}_0$.}
    \item The system response is observed over a 20-second window using the fourth-order SG dynamics for computational efficiency.
    \item Algorithm~\ref{alg:SGD} is employed to optimize the feedback gains based on the observed responses and the control objective in \eqref{eq:opt}, with ZOPG gradients.
\end{enumerate}}
We set the parameters as $r = 0.1$, $M=100$, $\eta = 10^{-4}$, \textcolor{black}{$J=15000$} and risk tolerance $c = 0.5$. Additionally, we set both $\mathbf{Q}$ and $\mathbf{R}$ as identity matrices for the LQR objective in \eqref{eq:opt}. The time step for both sensing and control is set to $\Delta t = 0.01~\textrm{s}$. 

\subsection{Training Result}
\begin{figure}[t]
	\centering
    \includegraphics[width=0.9\linewidth]{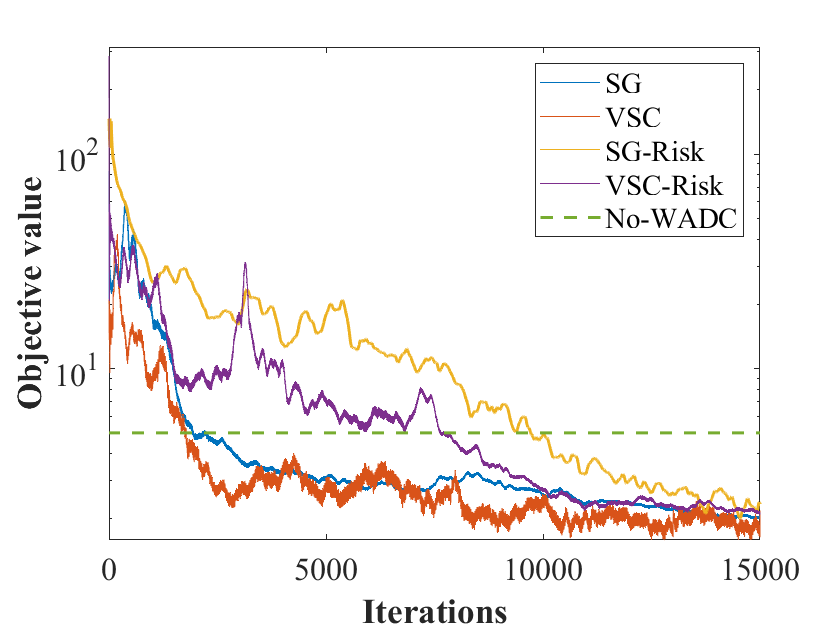}	
    \caption{Comparisons of the LQR trajectories during training for different WADC designs.} 
	\captionsetup{justification=centering}
	\label{fig:train}
\end{figure}

We first present the training results to verify the convergence of the proposed Algorithm~\ref{alg:SGD}. Fig.~\ref{fig:train} plots the log-scale objective trajectories for the four WADC designs, with the no-WADC cost objective as the baseline. Convergence has been observed for all trajectories, outperforming the baseline. Notably, \textit{SG-Risk} and \textit{VSC-Risk} exhibit higher fluctuation than \textit{SG} and \textit{VSC}. This is because the risk constraint would complicate the feasible region, and thus the search of optimal $\lambda$ leads to some oscillations in the trajectory. 
\textcolor{black}{Despite these fluctuations, the per‐iteration computational cost remains similar across all methods, and thus the computational time of training for \textit{SG‑Risk} and \textit{VSC‑Risk} is on the same order as the risk-neutral counterparts. Notably, all of them follow a model‐based SGDmax implementation that is known to be significantly faster than standard model‐free RL approaches.}

\begin{table}[t]
	\centering
	\caption{The WADC designs considered in numerical tests}
	\label{tb:cases}
	\begin{tabular}{ c  c  c }
	 \hline
		\textbf{} &\textbf{WADC} &\textbf{Risk constraint} \\ \hline
		\textit{SG} &SGs&Unconstrained \\
		\textit{SG-Risk} &SGs &Risk-constrained \\
        \textit{VSC} &SGs, VSCs &Unconstrained \\
		\textit{VSC-Risk} &SGs, VSCs &Risk-constrained \\
		 \hline
	\end{tabular}
\end{table}

\subsection{Testing Results}
The rest of the simulation results present the testing performance comparisons for the converged WADC policies obtained by training. 
\textcolor{black}{The testing phase evaluates the performance of the trained WADC policies under realistic conditions. As discussed in Section~\ref{sec:risk}, new scenarios are generated with different input disturbances and time-invariant link-specific delays, which are randomly sampled from a uniform distribution with bounds of 0.02s, 0.06s, and 0.10s, respectively. In this phase, the sixth-order SG dynamics are adopted to ensure an accurate representation of the system behavior \cite{generator}.} 

\begin{figure}[t]
	\centering
	\includegraphics[width=0.9\linewidth]{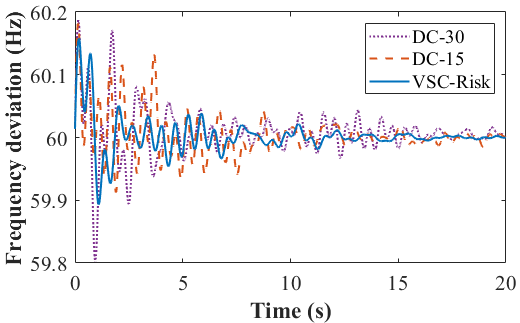}
	\caption{Frequency deviation comparison of our proposed VSC-Risk with conventional delay compensator (DC) based approaches.}
	\captionsetup{justification=centering}
	\label{fig:dc}
\end{figure}

{\KBC First, to verify the advantage of using risk-aware control, we compare the proposed \textit{VSC-Risk} with the conventional approach based on the delay compensator (DC);  see e.g., \cite{beiraghi}. The latter uses time tags of data packets to estimate the link delay and utilize it to reimburse the phase deviation and thus could incur estimation error. Accordingly, we compare with two DC settings, namely \textit{DC-15} and \textit{DC-30}, respectively having a maximum error of 15\% and 30\% on delay estimation. 
\textcolor{black}{Fig.~\ref{fig:dc} illustrates the frequency deviation of the bus~3 that generator~3 is connected to,
which represents the deviation of the actual frequency from the nominal 60Hz for each approach in a worst-case scenario with a maximum delay of 0.10s.} 
Our proposed \textit{VSC-Risk}, at no need of delay estimation, clearly outperforms \textit{DC-15} and \textit{DC-30}, as \textit{VSC-Risk} has the smallest frequency fluctuation. This comparison also verifies that the control performance of conventional DC-based WADC approach critically depends on the delay estimation accuracy, as \textit{DC-30} is observed to have increased the maximum frequency deviation and the time to reach steady-state as compared to \textit{DC-15},. 
Hence, this comparison has highlighted the advantage of using our proposed risk-aware WADC design in mitigating communication delays, thanks to eliminating the delay estimation step from conventional approaches.} 
%

\begin{figure}[t]
	\centering
	\includegraphics[width=0.9\linewidth]{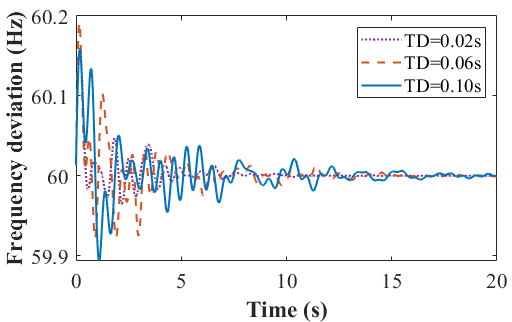}
	\caption{Frequency deviation of \textit{VSC-Risk} for different maximum delays.}
	\captionsetup{justification=centering}
	\label{fig:td}
\end{figure}

\textcolor{black}{Second, to illustrate the impact of communication delays, we select one specific scenario for each delay and plot the actual frequency deviation of bus~3 when using the \textit{VSC-Risk} design, as depicted in Fig.~\ref{fig:td}.} Each testing scenario has been selected to have the highest frequency deviation under each delay setting.
Clearly, the WADC performance degrades gradually with increasing delays, taking more time to dampen the oscillations. 

\begin{figure}[t]
	\centering
	\includegraphics[width=0.9\linewidth]{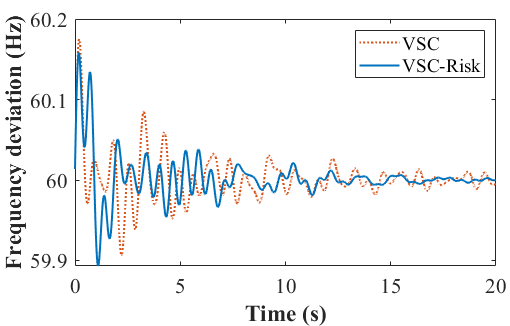}
	\caption{Comparison on the frequency deviation between \textit{VSC} and \textit{VSC-Risk}.}
	\captionsetup{justification=centering}
	\label{fig:risk}
\end{figure}

\begin{figure}[t]
	\centering
	\includegraphics[width=0.9\linewidth]{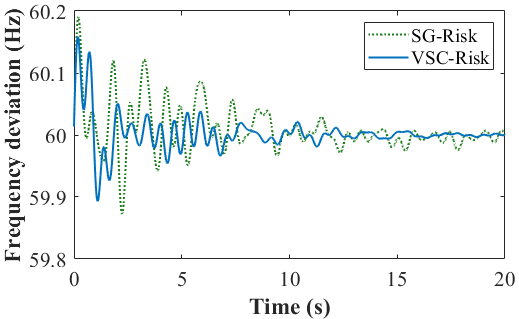}
	\caption{Comparison on the frequency deviation between \textit{SG-Risk} and \textit{VSC-Risk}.}
	\captionsetup{justification=centering}
	\label{fig:vsc}
\end{figure}

Moreover, to better demonstrate the effectiveness of integrating the risk constraint and VSCs into WADC, we compare the frequency deviation of the scenario with the highest frequency deviation under the setting of a maximum 0.10s delay.
First, Fig.~\ref{fig:risk} shows a frequency deviation comparison between \textit{VSC} and \textit{VSC-Risk} to demonstrate the effects of risk constraint. It has been observed that \textit{VSC-Risk} has more damping capability and reaches steady-state faster than \textit{VSC}. 
\textcolor{black}{This improvement is attributed to the risk constraint, as defined in \eqref{eq:opt}, which ensures that $R_c(\mathbf{K}) \leq c$. Specifically, \textit{VSC-Risk} achieves $R_c(\mathbf{K}) = 0.48$ when $c=0.5$, while \textit{VSC} results in a higher $R_c(\mathbf{K}) = 0.84$.}
This corroborates the usefulness of risk constraint in mitigating communication delays and maintaining the WADC performance. Second, Fig.~\ref{fig:vsc} compares \textit{SG-Risk} and \textit{VSC-Risk} to showcase the improvement of using VSCs for WADC. We can observe that the VSCs have provided additional actuation capabilities, leading to better damping performance.


\textcolor{black}{To better illustrate the damping performance of the proposed method, we conducted an eigenvalue analysis and introduced several modes along with their damping ratios derived from the eigenvalues of $(\mathbf{A}-\mathbf{B}\mathbf{K})$ in Table~\ref{tb:eigen}.}
Since our focus lies on the low-frequency range closely associated with WADC, we have introduced modes within the $[0.1, 2]$Hz range. By comparing (SG, VSC) with (SG-Risk, VSC-Risk), it is evident that incorporating the risk constraint enhances the damping performance, as we already discussed in Fig.~\ref{fig:risk}. Similarly, comparing (SG, SG-Risk) with (VSC, VSC-Risk), we observe that integrating VSCs improves damping performance, as illustrated in Fig.~\ref{fig:vsc}. This result emphasizes the effectiveness of our proposed method in enhancing worst-case performance with increased damping performance. \textcolor{black}{Despite using only three VSCs, our tests on the 68‑bus system could achieve an effective damping of the low‑frequency inter-area modes. Although the damping ofhigher‑frequency modes remains low, these local modes tend to be very lightly excited and have a minimal effect on wide-area oscillations.}


\begin{table}[t]
\centering
\caption{\textcolor{black}{Modes and damping ratios of eigenvalue analysis.}}
\label{tb:eigen}
\begin{tabular}{crrr}
\hline
\multicolumn{1}{l}{}      & \multicolumn{1}{c}{Eigenvalues}           & \multicolumn{1}{c}{Mode (Hz)} & \multicolumn{1}{c}{Damping ratio} \\ \hline
\multirow{5}{*}{SG}       & $-0.38 \pm j2.72$                       & 0.43                     & 0.14                        \\
                          & $-0.26 \pm j3,42$                       & 0.54                     & 0.08                        \\
                          & $-0.36 \pm j5.29$                       & 0.84                     & 0.06                        \\
                          & $-0.27 \pm j8.65$                       & 1.38                     & 0.03                        \\
                          & $-0.16 \pm j9.30$                       & 1.48                     & 0.02                        \\ \hline
\multirow{5}{*}{SG-Risk}  & $-0.43 \pm j2.71$                       & 0.43                     & 0.16                        \\
                          & $-0.38 \pm j3.39$                       & 0.54                     & 0.11                        \\
                          & $-0.35 \pm j5.30$                       & 0.84                     & 0.07                        \\
                          & $-0.28 \pm j8.45$                       & 1.35                     & 0.03                        \\
                          & $-0.19 \pm j9.13$                       & 1.45                     & 0.02                        \\ \hline
\multirow{5}{*}{VSC}      & $-0.55 \pm j2.49$                       & 0.40                     & 0.21                        \\
                          & $-0.48 \pm j3.29$                       & 0.52                     & 0.15                        \\
                          & $-0.37 \pm j5.26$                       & 0.84                     & 0.07                        \\
                          & $-0.27 \pm j8.65$                       & 1.38                     & 0.03                        \\
                          & $-0.15 \pm j9.30$                       & 1.48                     & 0.02                        \\ \hline
\multirow{5}{*}{VSC-Risk} & $-0.66  \pm  j2.47$                     & 0.39                     & 0.26                        \\
                          & $-0.55 \pm  j2.49$                      & 0.53                     & 0.16                        \\
                          & $-0.46 \pm j5.25$                       & 0.84                     & 0.09                        \\
                          & $-0.36 \pm j8.47$                       & 1.35                     & 0.04                        \\
                          & $-0.24 \pm j9.14$                       & 1.45                     & 0.03                        \\ \hline
\end{tabular}
\end{table}

\begin{figure}[t]
    \centering
    \subfigure[]{\label{fig:box2}\includegraphics[width=0.9\linewidth]{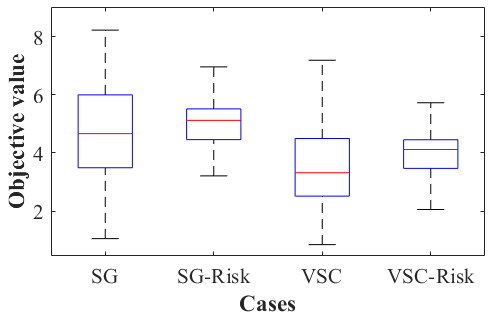}}	
    \subfigure[]{\label{fig:box6}\includegraphics[width=0.9\linewidth]{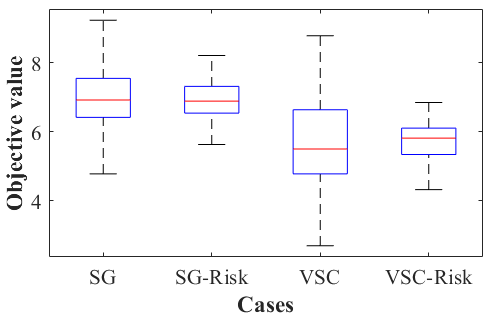}}
    \subfigure[]{\label{fig:box10}\includegraphics[width=0.9\linewidth]{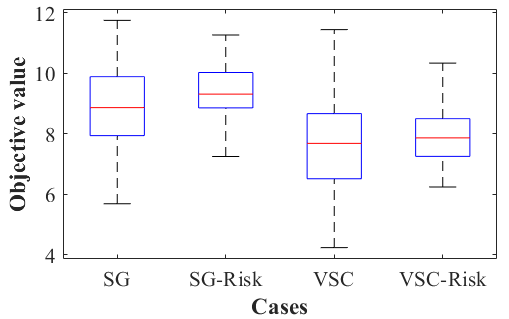}}
    \caption{Comparison of the objective values when the maximum delays are (a) 0.02s, (b) 0.06s, and (c) 0.10s.}
    \captionsetup{justification=centering}
    \label{fig:box}
\end{figure}

Furthermore, Fig.~\ref{fig:box} compares the statistical information of the LQR objective values over 100 testing scenarios.
Each subplot of Fig.~\ref{fig:box} has the box plots for all WADC designs with the median value, lower/upper quartiles and minimum/maximum, for each delay setting.
In general, the risk-constrained designs, both \textit{SG-Risk} and \textit{VSC-Risk}, have slightly increased the objective values on average, yet significantly reduced the variance and also the maximum of objective values. This result illustrates the effectiveness of using the risk constraint in mitigating the worst-case performance, thereby increasing the stability margin of power system operations. This comparison also verifies the improvements provided by the additional VSC resources, as \textit{VSC} and \textit{VSC-Risk} respectively outperform \textit{SG} and \textit{SG-Risk}.

\begin{figure}[t!]
	\centering
	\includegraphics[width=0.9\linewidth]{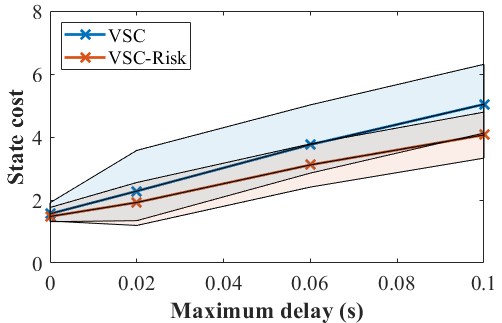}
	\caption{State costs of \textit{VSC} and \textit{VSC-Risk} under varying maximum delays.} 
	\captionsetup{justification=centering}
	\label{fig:scost}
\end{figure}

To further highlight the benefits of considering the risk constraint, we compare the state cost of \textit{VSC} and \textit{VSC-Risk} with increasing delays, as shown in Fig.~\ref{fig:scost}. The solid line represents the average value across 100 scenarios for each delay setting, while the blue and orange shaded areas indicate the state cost variations of \textit{VSC} and \textit{VSC-Risk}, respectively. 
It is evident that incorporating the risk constraint mitigates state cost increases at high delays and reduces variability more than the unconstrained cases. Thus, the risk constraint enhances the robustness of WADC design amid increasing delays in WAMS.

\begin{figure}[t]
	\centering
	\subfigure[]{\label{fig:c}\includegraphics[width=0.9\linewidth]{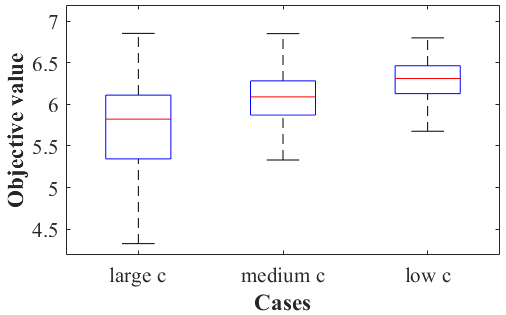}}	
	\subfigure[]{\label{fig:c_state}\includegraphics[width=0.9\linewidth]{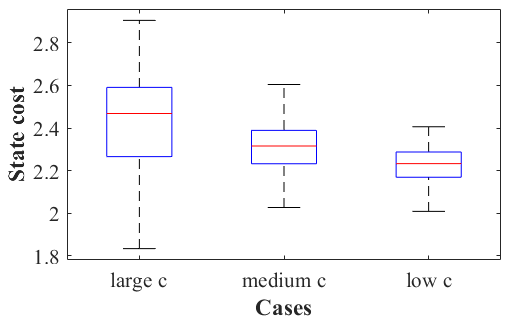}}
	\caption{(a) Objective values and (b) state costs for different risk tolerance parameters.}
	\captionsetup{justification=centering}
	\label{fig:diffc}
\end{figure}

Last, we investigate the impact of choosing the risk tolerance parameter $c$ with the highest delay setting. Fig.~\ref{fig:diffc} shows the box plots for the objective values and state costs for three different levels of $c$, with the same maximum delays of 0.10s.
In Fig.~\ref{fig:c}, a smaller value of $c$, which further limits the mean-variance risk, leads to a smaller objective value but higher variance. 
However, Fig.~\ref{fig:c_state} shows that both the state cost and the variance decrease with the $c$ value. This implies that further reducing the risk level in WADC could improve the performance in the state deviation, including frequency deviation, both in terms of the average value or the variance. However, this improvement may increase the overall LQR cost at the price of needing additional control efforts.

\subsection{\textcolor{black}{Robustness Analysis}} \label{sec:NT_RA}
\begin{figure}[t!]
	\centering
	\includegraphics[width=0.98\linewidth]{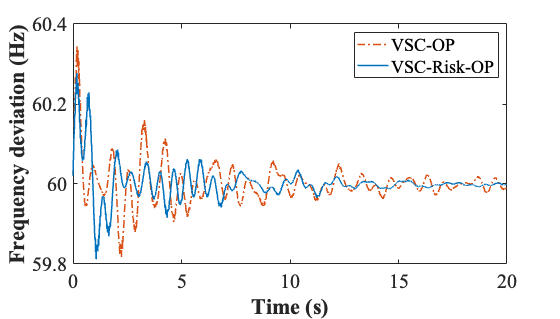}
	\caption{Comparison of frequency deviation between \textit{VSC-OP} and \textit{VSC-Risk-OP} at a 10\% perturbation of renewable outputs.} 
	\captionsetup{justification=centering}
	\label{fig:op}
\end{figure}

\textcolor{black}{
We validate the robustness to varying operating points by perturbing the renewable generation output of the IEEE 68‑bus system. We consider the RES perturbation by varying each VSC's power injection by up to ±10\% from its nominal value, and thus changing the   operating point (OP) and system model in  \eqref{eq:dyn}.
Fig.~\ref{fig:op} plots the frequency responses of both \textit{VSC-OP} and \textit{VSC‑Risk-OP}, by testing the respective pre-trained feedback gains on a selected 5\% perturbed OP scenario with a line outage. We have observed that the overall oscillation amplitude increases from that in Fig.~\ref{fig:risk} due to the additional OP perturbation. However, \textit{VSC‑Risk-OP} still achieves significantly faster damping and lower peak deviation than the risk-neutral \textit{VSC-OP}. This result corroborates the generalizability of our proposed risk-constrained controller in attenuating wide-area oscillations even when the operating conditions deviate from the training stage. 
}

\begin{figure}[t!]
	\centering
	\includegraphics[width=0.98\linewidth]{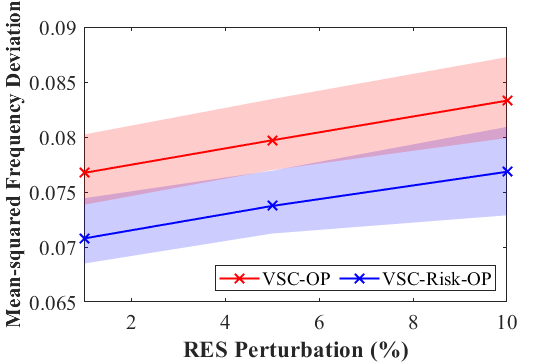}
	\caption{Comparison of MSFD values  between \textit{VSC-OP} and \textit{VSC-Risk-OP} over RES perturbation levels.} 
	\captionsetup{justification=centering}
	\label{fig:op_diff}
\end{figure}

\textcolor{black}{We further present some quantitative results for varying levels of RES perturbations in Fig.~\ref{fig:op_diff}. By perturbing the RES output by 1\%, 5\%, or 10\% with 1 to 3 randomly selected line outages, we generate 100 random OP scenarios. The  mean-squared frequency deviation (MSFD) is computed for each scenario over a total of $T$ time instances:  
\begin{align}
  \mathrm{MSFD} = \frac{1}{T}\sum_{t=0}^{T-1}\bigl(f_t - f_{\mathrm{ref}}\bigr)^{2},
\end{align}
by comparing the attained system frequency $f_t$  with the nominal $f_{\mathrm{ref}}$. Similar to Fig.~\ref{fig:scost}, the solid line in Fig.~\ref{fig:op_diff} denotes the average MSFD over the 100 scenarios, while the shaded regions indicate the ranges between the minimum and maximum MSFD values. Clearly, our \textit{VSC‑Risk‑OP} consistently reduces the MSFD values across all scenarios over the risk‑neutral \textit{VSC‑OP}. This comparison confirms the robustness of our risk‑constrained controller against a varying operating‑point with the topology changes, in reducing the worst-case frequency deviation and improving the system damping performance. 
}

\begin{figure}[t!]
	\centering
	\includegraphics[width=0.98\linewidth]{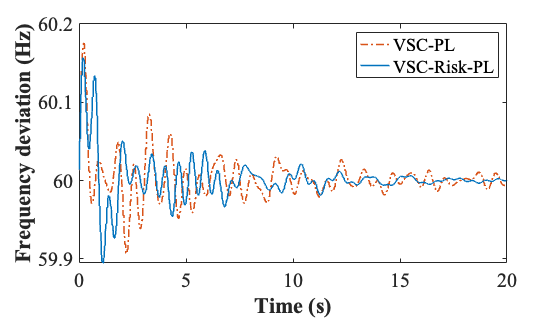}
	\caption{Comparison of frequency deviation between \textit{VSC-PL} and \textit{VSC-Risk-PL} at a 5\% probability of packet losses.} 
	\captionsetup{justification=centering}
	\label{fig:packet}
\end{figure}

\textcolor{black}{
In addition to RES perturbation, we have validated the generalizability of our risk-constrained WADC design under intermittent communication failures or packet losses (PL). We implement a simple slot‑based packet‑loss model under which at each time instance $t$, the packet carrying the PMU measurement is lost with probability $p\in[0,1]$. If a packet loss occurs, the controller holds the most recently received measurement of the state. Specifically, let $\gamma_t\in\{0,1\}$ denote an i.i.d.\ Bernoulli random variable with $\Pr(\gamma_t=0)=p$ and $\Pr(\gamma_t=1)=1-p$, and $\hat{\mathbf{x}}_t$ denote the measurement used by the controller at time $t$. Hence, we have $\hat{\mathbf{x}}_t = \mathbf{x}_{t}$ if $\gamma_t=1$; and otherwise, $\hat{\mathbf{x}}_t = \hat{\mathbf{x}}_{t-1}$.
%
Fig.~\ref{fig:packet} shows the resulting frequency deviation comparison between \textit{VSC-PL} and \textit{VSC-Risk-PL}, by testing the respective pre-trained feedback gains under this packet loss model with $p=5\%$. While the frequency deviation therein slightly increases from the loss-free counterparts in Fig.~\ref{fig:risk}, 
our
risk-constrained design remains superior over the risk-neutral one and  achieves a robust damping performance despite intermittent data loss.
}

\begin{figure}[t!]
	\centering
	\includegraphics[width=0.98\linewidth]{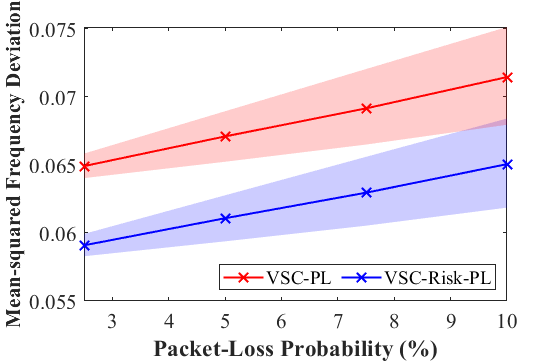}
	\caption{Comparison of MSFD values between \textit{VSC-PL} and \textit{VSC-Risk-PL} under varying packet-loss probability.} 
	\captionsetup{justification=centering}
	\label{fig:packet_diff}
\end{figure}

\textcolor{black}{Similarly, Fig.~\ref{fig:packet_diff} presents the MSFD ranges  under a varying packet‑loss probability $p$ value  from 2.5\% to 10\%. As $p$ increases, both \textit{VSC‑PL} and \textit{VSC‑Risk‑PL} exhibit growing MSFD values, demonstrating the effects of a higher frequency of packet losses by forcing the controller to use outdated data.
Nevertheless, our \textit{VSC‑Risk‑PL} maintains a consistent reduction in the MSFD values over the risk‑neutral \textit{VSC‑PL} across all $p$ values. This comparison again validates the robustness of our risk‑constrained controller against the adversary impact of packet losses, while confirming its generalizability to a range of uncertainty sources and various communication-induced perturbations.
}

In summary, our numerical tests validate the convergence of the proposed algorithm in solving the risk-constrained problem in \eqref{eq:opt}. Integration of VSCs has shown improved damping performance by providing additional actuation capabilities. Most importantly, the effectiveness of the risk constraint has been verified, especially in mitigating worst-case performance and improving system stability. \textcolor{black}{Lastly, the robustness analysis under both operating point perturbations and packet loss demonstrates that the proposed risk-aware WADC maintains superior damping performance even under significant uncertainties, highlighting its practical value for real-world deployment.
}


\section{Conclusions}\label{sec:CN}

\textcolor{black}{This paper designed a risk-constrained WADC approach that aims to address not only communication delays but also a broader range of cyber-physical uncertainties. Analysis of the LQR problem indicates that increased communication delays and cyber perturbation exacerbate state variability, and thus a mean–variance risk constraint is imposed on the state cost to bound this variability and improve worst‑case performance without requiring precise delay estimates. Using ZOPG-based gradient estimates, we developed an RL-based SGDmax algorithm and demonstrated its convergence and high computational efficiency.
Numerical tests on the IEEE 68-bus system have shown that our proposed design outperforms conventional delay compensator-based methods in the presence of delay estimation error, which also applies to other communication network issues such as link failures and cyber intrusions. In addition, we have verified that our risk-aware WADC design reduces the variability of total LQR cost, thereby improving system stability performance, especially in worst-case scenarios of oscillations and delays. The ability to mitigate delay-induced variability without requiring precise delay estimation highlights the robustness of our method, making it well-suited for modern power grid applications. 
}

\textcolor{black}{Future research directions include extensive baseline comparisons against model-free RL and robust LQR approaches, extensions to online RL for practical implementations, as well as investigating the scalability beyond the IEEE 68-bus benchmark. Furthermore, we aim to extend the framework to generalized grid control tasks for new resources, while incorporating physical constraints of voltage source converters, such as power limits, ramp rates, and inverter saturation, to enhance its realism and practicality. Finally, developing an online adaptive design will enable real-time policy adjustments to dynamic system changes, strengthening the applicability and robustness of the framework for future power grid applications.}

%
\bibliography{bibliography}
\bibliographystyle{IEEEtran}
\itemsep2pt





\end{document}